# CHIRAL STRUCTURES UNDER THE STANDARD OF ORTHOGONAL GROUPS

Salvatore Capozziello[a] and Alessandra Lattanzi[b]

[a]Dipartimento di Scienze Fisiche e INFN sez. di Napoli, Università di Napoli "Federico II", Complesso Universitario di Monte S. Angelo, Via Cinthia I-80126, Napoli, Italy.
E-mail: capozziello@na.infn.it.
[b]Dipartimento di Chimica, Università di Salerno, Via Ponte don Melillo, 84084, Fisciano, Italy.
E-mail: lattanzi@unisa.it.

*Abstract*: General features of microscopic and macroscopic chiral structures can be discussed under the standard of orthogonal group theory. Configuration space of systems, not physical space, is taken into account. This change of perspective allows to overcome traditional shortcomings related to true and false chirality, statistical realization of mirror images, classification of objects as "more" or "less" chiral. From this viewpoint, a chiral object is a physical system whose configurations are described by the O(N) algebra in an abstract N-dimensional space. A quantum mechanical interpretation is straightforward due to the fact that combinations of chiral states give rise to parity states which can be interpreted as energy eigenstates.

## 1. INTRODUCTION

Chirality is the underlying symmetry characterizing physical systems ranging from elementary particles to spiral galaxies.[1] Due to this fact, a huge amount of classification schemes for chiral objects has been proposed which, essentially, try to distinguish between "more" and "less" chiral systems,[2] between "true" or "false" chirality,[3] between fundamental and statistical chirality[4] and so on. Despite of all these efforts, few authors have tried to face the problem looking for a unifying approach able to overcome traditional classifications which take into account only single classes of systems (particles, molecules, macrostructures, etc.). The result of this way of thinking is that chirality is often considered only as a "geometric symmetry" with the final goal to find out only the most refined chiral index of a particular class of objects. On the other hand, chirality is also a "dynamical symmetry", not only related to rotations, reflections and inversions, but also to fundamental interactions and force laws governing the evolution of systems. Under this standard, chirality is a fundamental symmetry of nature, witnessing some process which has given rise to the enantiomeric modality by which we recognize almost all classes of systems. Besides, chirality is the footprint of the intrinsic asymmetry of physical objects which we observe in nature, ranging from left-handed neutrinos to L-amino acids, from D-sugars to trailing spiral galaxies. This points out that i) the expected achiral symmetry of nature has been broken by some dynamical process; ii) chirality is present at all scales, so then it is legitimate to look for a unifying scheme under which to enclose all these classes of systems and phenomena. However, the geometric symmetries are strictly related to dynamical symmetries and, following the "Erlangen program", according to Felix Klein,[5] every geometry and dynamcs of objects can be characterized by their own group of transformations. In this perspective, identifying the configuration space of a class of objects is the first step toward the transformations which such objects undergo and then toward the full dynamical description. This approach is successfully pursued in particle physics, where once identified the configuration space of systems (e.g. the configuration space of spins), the construction of the related Lie algebra allows to achieve the dynamics (e.g. the Hamiltonian of the interactions of the system). The problem is, however, more involved for mesoscopic, macroscopic or many-body systems but, if the specific configuration (or interaction) is recognized at least an effective description can be achieved. For chiral structures, a combination of spatial rotations and inversions seems the fundamental transformation characterizing systems at any scale and for any spatial dimension. Either simple or complex the structure is, the enantiomeric "modality" of any system cannot be absolutely recognized: it is realized by the identification of its mirror image recovered, in every case, by a combination of rotations and inversions which generates a non-superimposable system. Every (apparent or real) mechanism capable of giving rise to mirror system is a chiral transformation. This definition is completely independent of the effective physical realization of the enantiomer of the given system (particle, molecule, spiral galaxy, etc.), then a general chiral group can be, in principle, identified at any scale and for any physical dimension. In other words, a system which results "mirrored" after rotations and inversions is, in general, chiral or, using a physical concept, its Hamiltonian has to contain a chiral interaction term. However, it is well-known that the Hamiltonian function could not be easy to achieve for complex systems, but the presence of the above transformations is the probe of chiral dynamics.

In this paper, without saking for completeness, we want to face the problem of finding a group of transformations able to encompass a broad number of classes of chiral objects. From our viewpoint, this is the orthogonal $O(N)$ group, where N is the dimension of the configuration space. Due to this feature, the intrinsic chirality of an object is not related to the physical space where it lives, but to the abstract space of all possible configurations in which it can be realized. Furthermore, the total number of possible configurations is related to the generators of the $O(N)$ group and the approach results completely general, thus allowing to deal with spin-particles, tetrahedral molecules, helical structures or spiral galaxies under the same standard. In section 2, we discuss the general feature of a chiral object comparing some of the most popular definitions of chirality. It is shown that the configuration space plays a fundamental role in order to identify chiral transformations. Section 3 is devoted to the discussion of orthogonal groups $O(N)$. Their algebra emerges as the natural candidate to describe chiral transformations and chirality as a geometric symmetry. A general definition and a related theorem for chiral structures are given. In section 4, the cases of tetrahedrons, spin-particles and spiral galaxies are discussed as realizations of the above algebra. Section 5 is devoted to quantum mechanical considerations, to the discussion of chiral transformations with respect to parity and quantum states. The goal is to show that orthogonal groups describe chirality also at fundamental level and then are related to dynamical symmetries of the systems. Conclusions are drawn in section 6.

## 2. LOOKING FOR A CHIRAL GROUP

*C'est la dissymétrie qui creé le phénomène.*[6] This famous sentence by Curie can be considered the essence of chirality which emerges as the property which differentiates an object from its mirror image. Such a "dissymetry" is related to some fundamental symmetry breaking mechanism (e.g. CP violation)[7] and manifests itself at any scale ranging from objects like neutrinos up to huge systems as spiral galaxies.[8] In particle physics, chiral symmetry is broken and, for example, for small quark masses, we can view the π meson as the Goldstone boson for the broken chiral symmetry. The fact that π meson has a light mass is a good indicator of the fact that chirality is an approximate symmetry, or, in other words, a dissymmetry.[9] In general, chiral symmetry of particles can be described by $SU(N) \otimes SU(N)$ transformations, where N is related to the specific interaction (N=2 for Dirac spin-particles, N=3 for Quantum Chromodynamics, etc.). Clearly, a fundamental role is played by N, the dimension of the abstract configuration space where the interaction is described. From this point of view, chirality is not a feature of the physical space where particles live, but a property of the group of quantum mechanical transformations they undergo. This concept will be extensively discussed in the conclusions considering also that the distinction between chiral and achiral objects can be directly related to the dimension of the embedding space.[10]

On the other hand, in chemistry, chirality is experimentally manifested by "pseudoscalar measurements" related to optical rotation and circular dichroism. Due to this fact, it is often considered only a genuine spatial property of molecules and then described in a completely different fashion with respect to its particle counterpart, although the issue that particle and molecular chirality has the same fundamental origin has been often supported.[10] Considering molecular chirality as a geometric symmetry has led to the classification of molecules as achiral compounds, diastereoisomers, enantiomers.[11] In other words, chemical systems can be divided into sets containing chiral objects: if the chiral objects are handed (i.e. shoe-like), conventionally we have "left" and "right" objects; if the chiral objects are non-handed (i.e. potato-like), they are neither identical with nor a mirror image of a non-handed object.[12] The situation becomes more involved for complex objects as octahedrons or helical structures as nucleic acids (DNA) and polymers. People are searching for classification schemes able to point out the chirality degree (chirality indexes). In this respect, it has been proposed to measure the minimal distance between a chiral compound and a reference achiral structure, which is the basic concept of the "continuous chirality measure" methodology.[13] Quantitative correlations between these chirality measurements and chemical, physical parameters can be established. Moreover, chirality is connected to the inability to make a structure coincident with a statistical realization of its mirror image;[4] the probe-dependent measurement of this inability is the chirality content of the structure. This last definition results operative for large random supramolecular structures as spiral bacterial colonies, spiral hurricane cloud formations and spiral galaxies. However, even if this state of art works to describe single classes of objects, it is unsatisfactory from the viewpoint of symmetry and conservation laws, since it seems that a different kind of chirality descriptor is necessary at different scales. Conversely, the nature of chirality would depend on the size of objects and this is not reliable in a unitary picture of science. For example, taking into account fundamental interactions as gravitation or electromagnetism, a suitable working hypothesis is that they act in the same way from microscopic to macroscopic scales. This hypothesis is supported by experiments and observations, so the intrinsic unitarity of physics (covariance and invariance of physical laws) is a standard paradigm. This result should be achieved also for chirality at any scale, since it is a general feature of nature. With these considerations in mind, it is possible to find out a group of transformations whose algebra works at any scale and for any chiral object? Such a group can be related to the configurations that the object can assume without taking into account the physical space where the object is embedded? Answering these questions would represent a non-trivial step to deal with chirality as a dynamical symmetry and then understanding its features independently of the physical size and spatial dimension of chiral objects. In the next section, we show that orthogonal group $O(N)$ present several features useful to fully achieve this program at every physical scale. Interesting examples will be discussed in section 4.

## 3. ORTHOGONAL GROUPS AND THEIR RELATIONS TO CHIRALITY

Starting from Kelvin definition: *Any geometric figure, or groups of points is chiral, and it has chirality, if its image in a plane mirror, ideally realized, cannot be brought to coincide with itself.*[15] The corresponding definition of absence of chirality can be expressed saying that a structure and its mirror image are superimposable by rotation or any motion preserving the structure. It is clear that, independently of the object size, a chiral transformation has to be constituted by a combination of rotations and inversions. This statement works for configurations of objects considered in the physical space (e.g. molecules) or in the abstract space of spins (e.g. particles). Furthermore, the size and the structure of the object has to be conserved in the "mirroring" transformation.[16] This means that the transformation has to be unitary or, in other words, distances in N dimensions have to be preserved. Specifically, the distance from the origin to the point $x^i$, by the Pythagorean theorem, is given by         . Therefore, $x^i x^i$ is an invariant and the N-dimensional rotation is defined by

$$\quad (1)$$

The number of independent elements in each member of $O(N)$ is $N^2$ minus the number of constraints arising from the orthogonality condition, that is:

$$\quad (2)$$

This is the number of independent antisymmetric NxN matrices, that is, we can parametrize the independent components within $O(N)$ by either orthogonal matrices or by exponentiating antisymmetric ones. Any orthogonal matrix can, thereby, be parametrized as

$$ \tag{3} $$

where $\tau^j$ are linearly independent, antisymmetric matrices with purely imaginary elements. They are called the "generators" of the group, and $\theta^j$ are the rotation angles or the "parameters" of the group. Finding representations of $O(N)$ is complicated, however, the Baker-Campbell-Hausdorff theorem can be used. It states that

$$ \tag{4} $$

where [A,B] is a commutator.[17] If $e^A$ and $e^B$ are close together, then these elements form a group as long as the commutators of A and B form an algebra. Without entering into details of group theory, we want to show that $O(N)$ groups are sufficient to represent any chiral transformation in the configuration space of a given system. Let us consider, for the sake of simplicity, the $O(2)$ case and the fact that "enantiomer" means that a physical system can be present in two modalities with non-superimposable mirror images. Being N=2, the number of generators of transformations is 2(2-1)/2=1, corresponding to a rotation angle. Any rotation in the plane can be written as

$$ \tag{5} $$

which is a particular case of (1). Because the group is orthogonal, it follows that       and then

$$ \tag{6} $$

which means that the inverse of an orthogonal matrix is its transposition, that is $O^{-1} \equiv O^{T}$. Considering the determinant of both sides of eq. (6), we get

$$\tag{7}$$

that is the determinant of *O* can be ±1. Considering the subset with det(*O*)=1, we have the subgroup *SO*(2) of special orthogonal matrices in two dimensions describing 2D rotations. The subset of *O*(2), given by det(*O*)=-1, does not constitute a group. It consists of elements of *SO*(2) times the matrix

$$\tag{8}$$

and then gives rise to the transformation

$$\tag{9}$$

which takes a plane and maps it into its mirror image. In summary, *O*(2) represents the group of all possible rotations and inversions in a 2D-space independently of the nature of the space and the object described in it. Any 2D-chiral transformation, where an object is "mirrored" in itself can be represented by a one-parameter continuous rotation plus a discrete reflection. If the result of such an operation gives rise to the enantiomer of a given system, the transformation has determinant det(*O*)=-1; if the "mirrored" object results identical to the starting one, det(*O*)=1. This kind of transformations can be represented also by complex numbers. In this case, if          is a complex number (a vector in the complex plane), it can be transformed as

$$\tag{10}$$

where *U*(  ) is a complex unitary matrix

$$\tag{11}$$

where    is the Hermitean conjugate operator of *U*. The set of all one-dimensional unitary matrices          is the group *U*(1) where the multipication law

$$\tag{12}$$

holds. This multlipication law is the same of *O*(2) even though this construction is based on the one-dimensional space of complex number. Then the correspondence

$$\tag{13}$$

is straightforward: this means that two real numbers transformed by *O*(2) can be combined into a single complex number transformed by *U*(1). In other words, a transformation given by a rotation and an inversion in a real space is equivalent to a complex conjugation in a complex space. Again, a chiral transformation can be reduced to a complex conjugation. The previous considerations can be extended in a **Theorem**: *chiral transformations in any N-dimensional space are given by the transformations of the corresponding O(N) algebra. The number of independent generators is [N(N-1)]/2 (i.e. the number of independent parameters), the number of possible configurations of the system is N! where ½(N!) belongs to an enantiomer and ½(N!) to its opposite mirror image.*

This theorem can be immediately translated into a complex number representation, if we adopt *U*(N) (or *SU*(N)) groups. In the following section, we shall give some peculiar examples of chiral systems which can be represented under the unifying view of orthogonal groups independently of the physical space in which they are embedded. However, it is worth nothing that an object is chiral or achiral depending on the dimension N of the configuration space: it is chiral only in the lowest dimensional space where it is embeddable.[10]

## 4. EXAMPLES OF CHIRAL STRUCTURES

As a first application of the above considerations, let us take into account tetrahedral molecules[18] whose configuration space can be easily achieved by the well-known Fischer projections.[12] This representation is based on some simple empirical rules whose aim is to achieve planar projections of the molecule: the atoms pointing sideways must project forward in the model, those pointing up and down in the projection must extend toward the rear. In the following picture, these rules are illustrated for the lactic acid as a model compound (Figure 1).

**Figure 1**: Fundamental rules to handle Fischer projections.

Starting from a fundamental projection, rotation of 90° are not allowed, while 180° rotations give correct results. The interchange of any two groups gives rise to the mirror image of a given enantiomer. The chemical groups are indicated by numbers running from 1 to 4. Without considering their priorities according to Cahn-Ingold-Prelog (CIP) rules,[19] we can set: $OH=1$, $CO_2H=2$, $H=3$, $CH_3=4$. There are $4!=24$ permutations of 4 ligands and 12 of these correspond to the (+)-enantiomer (Fig. 2) and the other 12 graphs to the (-)-enantiomer (Fig. 3).

**Figure 2**: Fischer projections of (*S*)-(+)-lactic acid.

**Figure 3**: Fischer projections of (*S*)-(-)-lactic acid.

Permutations in Fig. 2 can be achieved either permuting groups of 3 bonds or by turning the projections by 180°. Permutations in Fig. 3 are obtained by those in Fig. 2 interchanging two chemical groups. Considerations of section 3 can be immediately applied to this case. The tetrahedral molecule is represented as a column vector $\mathcal{M}$ in a 4D space, i.e.

$$\mathcal{M} = \begin{pmatrix} \Psi_1 \\ \Psi_2 \\ \Psi_3 \\ \Psi_4 \end{pmatrix} \qquad (14)$$

The 24 Fischer projections in Figs.(2) and (3) can be realized from the fundamental one through an $O(4)$ algebra whose elements are the 4x4 matrix-operators in Tables I and II acting on the fundamental projection.

**Table 1**: (+)-Enantiomer.

**Table 2**: (-)-Enantiomer.

Clearly, the matrix-operators in Table 1 are rotations with         , while operators in Table 2 are inversions with          . The generalization to chains of molecules with n stereogenic centres is straightforward.[20] Through this approach, a molecular "aufbau" method, useful for classification of enantiomers, diastereoisomers and achiral molecules can be achieved. As final assessment, the chiral algebra is independent of physical 3D space of the molecule while the chiral features emerge thanks to the transformation properties in the 4D configuration space.

On a completely different ground, it is possible to show that the *O*(4) chiral algebra works also for spin-particles. In this case, regardless to the size of the systems, particles and molecules undergo exactly the same transformations. For example, a Dirac spin-vector (describing electrons and positrons with the same mass, opposite charges and obeying the same Dirac equation) can be described as a relativistic 4-vector field

$$= \begin{pmatrix} \Psi_1 \\ \Psi_2 \\ \Psi_3 \\ \Psi_4 \end{pmatrix} \qquad (15)$$

representing the spin-up-down states of an electron-positron multiplet.[21] Any spin transformation is a member of the 2D special unitary group *SU*(2) and the isomorphism

$$\qquad (16)$$

holds for the whole multiplet (15). A chiral transformation, combining the 4-momentum and the helicity, is easily achieved by the pseudoscalar chiral operator

$$\qquad (17)$$

where **I** is the 2x2 unit matrix and the 4x4 matrices $\gamma^i$ are the Dirac matrices

$$, \qquad . \tag{18}$$

The 2x2 matrices

$$\tag{19}$$

constitute the *SU*(2) Pauli algebra. It is interesting to observe that the matrix γ⁵ corresponds to the above matrix       in Table 1. The 4x4 matrices $\gamma^0$, $\gamma^i$, $\gamma^5$, plus the 4x4 unit matrix constitute a chiral algebra equivalent to the previous *O*(4) algebra of tetrahedral molecules thanks to the isomorphism (16). In the previous case, we have used a real representation of the algebra, in this case, the complex representation is due to the fact that charge conjugation of Dirac spinors is represented by complex matrices.[7] The corresponding "enantiomers" can be identified in the 4D configuration space as a left-handed fermion, whose spin is anti-parallel to its momentum vector (negative helicity) and a right-handed fermion, whose spin is parallel to its momentum vector (positive helicity) as depicted in Figure 4.

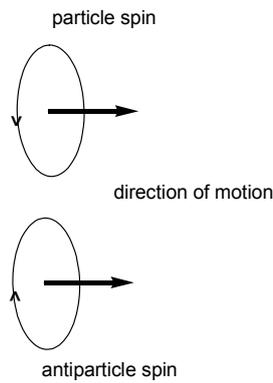

**Figure 4**: Chirality of Dirac spin particles.

We remark again that such a form of chirality can be identified only in the 4D abstract space of configurations and not in physical space. If we take into account a different space of configurations, such a chiral feature is not present.[10]

The orthogonal group approach to chirality works also for extremely macroscopic systems once a configuration space is correctly identified. For example, it is easy to show that the spiral structure of galaxies is intrinsically chiral.[22] In fact, trailing and leading arms of spiral galaxies require the kinematical determination of the galaxy plane orientation with respect to our line of sight, so that radial velocity can be interpreted in terms of the direction of the galaxy rotation.[23] Taking into account the recession velocity, given by the Hubble relation          , where $H_0$ is the Hubble constant and "d" the distance of the galaxy from the observer, we can assign helicity moving along the arms of the galaxy toward the center (Figure 5).

**Figure 5.** Trailing and leading modes of spiral galaxies.

Trailing galaxies are left-helical while leading galaxies are right-helical. The two systems result non-superimposable with respect to reflections within the plane. In other words, the enantiomers in 2D, are the trailing arm galaxy rotating in one sense and the leading arm galaxy rotating in the opposite sense. This means that rotation of the system has to be considered with respect to recession velocity. In summary, the configuration space of a spiral galaxy is globally assigned by a rotation and a reflection so that *O*(2) orthogonal group is suitable to describe galaxies as chiral structures. Even in this case, information on the size and the physical space in which the system lives is not required. Several other chiral (mesoscopic or macroscopic) systems can be described under the standard of orthogonal groups. For example, spiral hurricane cloud formations can be dealt as *O*(2)-trailing or leading systems. The "enantiomers" are represented by their clockwise or anti-clockwise rotation modality with respect to the Earth equator line.

Short helical systems as helicenes and long analogs as DNA, proteins and synthetic polymers can be globally described as chiral systems considering the transformations of the $O(2)\otimes$ -composed group, including rotations, inversions and translations. In conclusion, any chiral system at every scale can be dealt under the standard of orthogonal groups once the global transformations which it can undergo are correctly identified. This operation allows to construct the configuration space of the system, i.e. the dimensionality N, the number of possible configurations N!, the number N(N-1)/2 of generators of transformations which is the number of independent parameters and the procedure is completely independent of the physical space where the system lives.

## 5. QUANTUM MECHANICAL CONSIDERATIONS

From the fundamental physics viewpoint, geometric symmetries are always related to dynamical symmetries. The first ones refer to the geometric structures of the systems and includes rotations, reflections, inversions and translations. The second ones relate to the particular form of the interactions among the different parts of the system. In a certain sense, the second ones yield the first ones, which are the easiest to be identified. Conversely, identifying the whole group of transformations which a system undergo is the first step toward the full dynamical description of the system. In the specific case of chiral structures, if they are described by $O(N)$ groups, this means that related dynamics has to be associated to orthogonal transformations and the Hamiltonian operator of the systems has to commute with orthogonal operators. However, dynamics and geometric symmetries have to be considered into the configuration space of the system and then into the phase space. In general, when studying the chiral dynamics, we can suppose that chiral states interconvert between the left- and right-handed states, which, therefore do not have definite parity. In other words, we can suppose that any racemic parity-defined state is given by the superposition of two enantiomers of a given chiral system. We can take into account the set of operator      , $\chi_k$ and $\bar{\chi}_k$ where      is the total

Hamiltonian operator of the degenerate generalized "isomer" consisting of even and odd part

$$\quad (20)$$

$\chi_k$, $\bar{\chi}_k$ are chiral operators giving rise, respectively, to rotations and inversions in a given N-dimensional configuration space and k is an index running from 1 to (1/2)N!. For the sake of simplicity, let us ignore parity violations. In this picture,[24] energy eigenstates are superpositions of handed states, i.e.

$$\quad (21)$$

It is worth noting that the chiral system has not been specified, so the discussion is completely general. Immediately, the relation

$$\quad (22)$$

holds for the parity operator whose eigenvalues are . Dropping the indices and considering any chiral state of the system in the configuration space, it is

$$; \quad (23)$$

$$; \quad (24)$$

The $\bar{\chi}_k$ operators interconvert the two handed states, while $\chi_k$ are pure rotations leaving unchanged the handness of the state. Let us consider now the energy eigenstates (21) which are also parity eigenstates thanks to (22). We have

$$\quad (25)$$

and

$$\quad (26)$$

That is the N! $\chi_k$ and $\bar{\chi}_k$ operators of the *O*(N) algebra leave unchanged the parity of energy states (i.e. parity is conserved). This result is completely general and works for microscopic, mesoscopic and macroscopic chiral systems. We have used Dirac notation in order to stress that the approach is completely independent of physical space in which the system is set. However, we should consider processes as P or CP-violation in order to achieve a detailed dynamics capable of matching experimental data, but our aim here is to show that chirality, under the standard of *O*(N) groups, can be dealt also as a dynamical symmetry other than a geometrical one.

## 6. CONCLUSIONS

In this paper, we have discussed an approach by which it is possible to describe chiral structures regardless their size, but considering only the transformations which they undergo in the configuration space. This way of thinking leads to take into account orthogonal groups *O*(N) which encompass all unitary transformations, rotations and inversions, by which it is possible to establish if the (real or virtual) mirror image of a system is superimposable or not with respect to the given one. This approach is general to define chirality of systems and it does not depend on the size and the fact that we are considering either quantum or classical systems. In this way, the chirality of completely different systems as particles, molecules or spiral galaxies can be described under the same standard. The crucial points are the identification of the configuration space dimensions, the independent parameters or generators of transformations, the suitable representation of the unitary group which can be achieved by real or complex numbers. The number of rotation and inversion operators is related to the dimensionality of the space and to the number of configurations of the system. It is possible to show that such a geometric chiral symmetry is also a dynamical symmetry related to the energy eigenstates of the system. However, this statement rigorously holds if symmetry violations (as CP or P-violation) are not considered, but it is extremely general since it does not depend on the intrinsic nature of the system. In other words, we only need that a physical system could be globally described as an isomer presenting two enantiomeric states. Dynamical symmetry is guaranteed by the fact that combinations of chiral states give rise to definite parity eigenstates which are also energy eigenstates. Such an approach has to be furtherly refined, first of all in relation to the symmetry breaking processes but, in any case, it could be a further step to frame microscopic and macroscopic chiral systems into the same theoretical scheme. However, some considerations on the effective meaning of chirality with respect to the embedding space are in order at this point. It can be formally proven that any chiral object embedded in an N-dimensional space is achiral when embedded in any space of dimension higher than N. In other words, an object is chiral only in the lowest dimensional space where it is embeddable.[10] This property shows that chirality is a genuine geometric feature which emerges if configuration space (and in particular its dimensionality) is chosen in a suitable way. Furthermore, the configuration space of N-particle systems can show aspects directly related to the measure of chirality, as shown in ref. 25, and then the approach presented in this paper is far from being conclusive. Finally, in order to achieve a self-consistent theory, the problem of chirality measure for simple and complex systems, has to be faced. In particular, our considerations have to be related to the metric properties of chirality and to the mechanisms capable of preserving chirality. For a comprehensive discussion of these topics see ref. 26.